# Energy Efficient Virtual Machine Services Placement in Cloud-Fog Architecture

Hatem A. Alharbi, Taisir E.H. Elgorashi and Jaafar M.H. Elmirghani
School of Electronic and Electrical Engineering University of Leeds, LS2 9JT, United Kingdom

*Abstract*— The proliferation in data volume and processing requests calls for a new breed of on-demand computing. Fog computing is proposed to address the limitations of cloud computing by extending processing and storage resources to the edge of the network. Cloud and fog computing employ virtual machines (VMs) for efficient resource utilization. In order to optimize the virtual environment, VMs can be migrated or replicated over geo-distributed physical machines for load balancing and energy efficiency. In this work, we investigate the offloading of VM services from the cloud to the fog considering the British Telecom (BT) network topology. The analysis addresses the impact of different factors including the VM workload and the proximity of fog nodes to users considering the data rate of state-of-the-art applications. The result show that the optimum placement of VMs significantly decreases the total power consumption by up to 75% compared to a single cloud placement.

## I. INTRODUCTION

Cloud computing is an immensely successful on-demand computing paradigm that provides ubiquitous access to a shared pool of compute, storage, and communication resources to a large set of geographically distributed users. According to Cisco [1], in 2017, the total clouds traffic was 75% of all United Kingdom (UK) Internet traffic. A further growth is projected within the approaching years as the total cloud computing traffic is expected to be 91% of the total traffic by 2022. This proliferation in data volume and processing requests raise the needs for a new breed of on-demand computing. Fog computing [2] is proposed by academia and industry in order to complement the cloud by extending processing, networking and storage resources of clouds to the edge of the network bringing them closer to the users.

Clouds and fog employ virtual machines (VMs) for efficient resource utilization [3], [4]. In order to achieve the most of the efficient environment, VMs can be migrated or replicated over geo-distributed physical machines in order to achieve different features such as load balancing and energy efficiency [6] or due to machines maintenance. The workload of VMs has been intensively investigated in the in the literature. They found that the relationship between the VM workload and the number of user typically follows a linear profile as shown in Fig. 1. The minimum VM workload required to serve users varies from as low as 1% in data-intensive applications to 40% in CPU-intensive applications [7] - [9].

Research efforts have focused on using fog computing to address the challenges facing cloud computing. In terms of QoS, a mathematical model of fog computing was developed in [10] to investigate the possibility of reducing the latency of IoT applications. In [11], the authors considered improving websites performance by connecting users to the Internet via fog servers. A number of papers in the literature investigated the energy efficiency of fog architecture. The authors in [12] built a theoretical model of the fog computing architecture and compared it with conventional cloud computing. The authors in [13] compared the energy consumption of applications run on a centralized cloud with their energy consumption when hosted in the fog. Their results showed that the energy efficiency of fog computing is determined by a number of factors including; the energy efficiency of the access network, the power profile of fog servers and the application type in terms of the number of downloads, the number of updates and the amount of data pre-loading.

Significant research efforts in the literature have focused on cloud computing and communication networks energy efficiency [14]-[34]. In this paper, we optimize the network and processing energy efficiency by developing a framework for VMs placement over cloud-fog architecture. We build a mixed integer linear programming (MILP) model to minimize the total energy consumption by optimizing the placement of VMs based on multiple factors including the VM workload, the workload versus number of users profile and the proximity of fog nodes to users considering the data rate of state of art applications. The remainder of this paper is organized as follows. Section II introduces the proposed Cloud-Fog Architecture. In section III, we present and discuss the VMs placement model and results. The paper is concluded in Section IV.

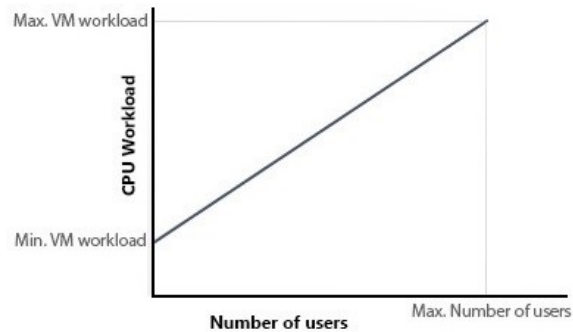

Figure 1: Linear relationship between VM workload and number of users.

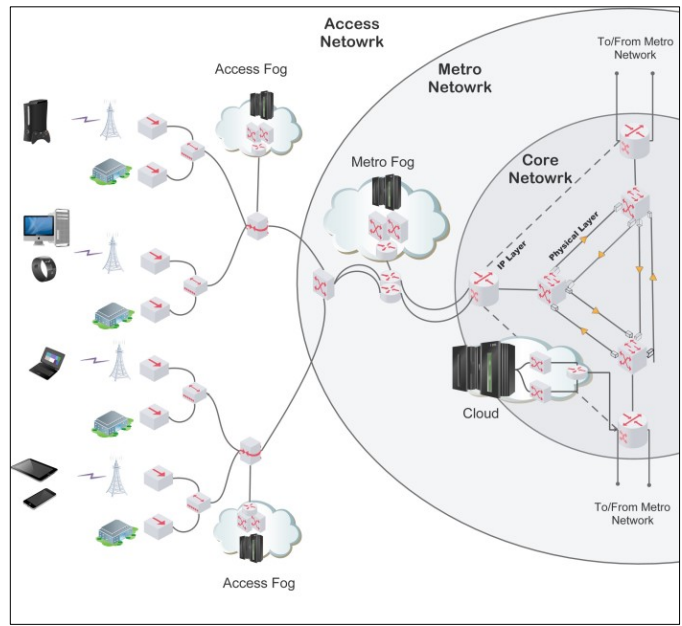

Figure 2: Cloud-Fog architecture.

## II. CLOUD-FOG ARCHITECTURE

In order to determine the optimal placement of different VMs types, we built a MILP model considering different tiers of cloud and fog resources. As shown in Fig. 2, a typical telecom network consists of three layers; core, metro and access network. The core network represents the backbone infrastructure of any telecom network as it interconnects major cities. IP over WDM [35] technology is widely deployed in core network due to its ability to provide high capacity and scalability. Each core node is connected to a metro network. The metro network provides direct connectivity between residential users and the core network node. The access network represents the last mile of the telecom network and connects the telecom office to end-users. Passive optical networks (PON) are the main technology deployed recently in access networks. It has two active components; Optical Line Terminal (OLT), located at the telecom office, and Optical Network Unit (ONU), located close to end-users.

In cloud-fog architecture, clouds provide a centralized computing solution to end users. Cloud data centers are connected to IP routers in a core node. Fog computing extends clouds into the network edge. In our work, we consider fog nodes deployed in both the metro and access networks. Metro fog nodes are connected to metro edge routers and can serve users of different access networks. The access fog nodes which are connected to the OLTs, provide a smaller scale coverage by serving only users in their access network.

## III. VMs PLACEMENT MODEL RESULTS

We investigated the optimal VMs placement over the BT network topology [36] as a core network example (illustrated in Fig. 3). The BT core network topology consists of 20 nodes and 68 bidirectional links. In the considered architecture each core node is connected to two XGPON access networks through a metro network consisting of a single Ethernet switch and two edge routers.

The total capacity of each XGPON OLT is 1280 Gbps [37]. According to Cisco Visual Network Index (VNI) [38], in 2017, the average broadband speed in the United Kingdom (UK) was 33.7 Mbps. Consequently, we assumed that each OLT is able to serve ~38k connections (or users). Also, Cisco VNI reported that in 2017, 75% of all Internet traffic had crossed clouds. So, in each OLT, we consider a scenario in which 28.5k users out of the 38k users access applications hosted in VMs. SimilarWeb analytics tools [39] reports that the top 300 applications / websites have a share of 66% of all clouds traffic. Accordingly, we consider 18.8k users in each XGPON to access the 300 applications / websites hosted in VMs. The popularity of these applications / websites VMs is considered to follow a Zipf distribution [40]. To simplify our analysis, we grouped VMs of similar popularity into 6 groups with the following average popularity; 16%, 5%, 2%, 1%, 0.5% and 0.05%. The number of VMs in each popularity group is 1, 4, 5, 9, 43 and 238, respectively. Each VM is considered to require 50% of the CPU's server capacity in order to serve 800 users [41]. The VM workload vs number of users served by the VM is considered to follow a linear profile with a minimum CPU usage of 1% of the total server CPU capacity based on the CPU requirements for state of the art applications [8]. The users are considered to access the VMs with one of the following data rates; 1 Mbps, 10 Mbps or 25 Mbps. Such data rates represent the recommended download speed to access the content of the state of the art applications, e.g. 1 Mbps for light web browsing [42], 10 Mbps for application processing high-definition video quality [43] and 25 Mbps for application processing ultra-high video quality [44]. Tables I and II show the input parameters of the cloud-fog architecture model. Note that, we considered on-off power consumption profile for the different components in the architecture.

The optimized VMs placement over the cloud-fog architecture, referred to as *Optimized clouds and fogs placements (OC&F)* approach. It is compared to the *Optimized clouds* (OC) approach where VMs are optimally placed in clouds distributed over the core network. It is also compared to *the single clouds* (SC) where the VMs are placed in node 6 (City of London). Node 6 is selected to host the cloud in SC approach as major cloud operators base their central cloud in the UK in London (e.g. Microsoft Azure [45], Amazon AWS [46] and Google Cloud [47]).

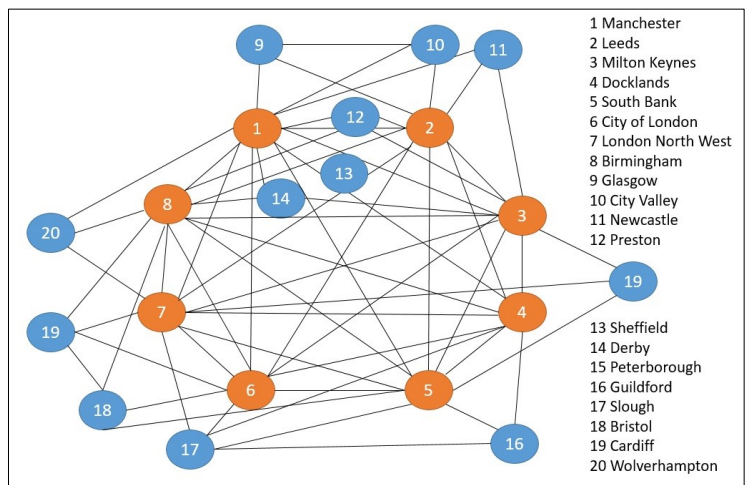

Figure 3: BT core network topology.

TABLE I
NETWORK INPUT PARAMETERS OF THE MODEL

| | |
|---|---|
| 40 Gbps router port power consumption | 638 Watt [48] |
| 40 Gbps transponder power consumption | 129 W [49] |
| 40 Gbps regenerator power consumption | 114 Watt, reach 2000 km [50] |
| EDFA power consumption | 11 Watt [51] |
| Optical switch power consumption | 85 W [52] |
| Number of wavelengths in a fiber | 32 [53] |
| Bit rate of each wavelength | 40 Gbps [53] |
| Span distance between two EDFAs | 80 km [51] |
| Network power usage effectiveness | 1.5 [6] |
| Aggregation router redundancy | 2 |
| 40 Gbps aggregation router port power consumption | 30 Watt [54] |
| 600 Gbps metro ethernet switch power consumption | 470 W [55] |
| Number of PON network in a node | 2 |
| Number of network users | 800 concurrent users per VM. |
| Number of ONU devices in a PON network | 512 |
| Power consumption of ONU device | 5 W [55] |
| OLT Power consumption | 1842 W [37] |

TABLE II
CLOUDS AND FOG INPUT PARAMETERS OF THE MODEL

| | |
|---|---|
| User download rate | {1, 10 and 25 Mbps} |
| 600 Gbps cloud and metro fog switch power consumption | 470 W [55] |
| 240 Gbps access fog switch power consumption | 210 W [55] |
| Cloud and fog switch redundancy | 2 |
| 40 Gbps cloud router port power consumption | 30 Watt [54] |
| 40 Gbps metro and access fog router port power consumption | 13 Watt [54] |
| Best practice cloud power usage effectiveness | 1.3 [57] |
| Best practice metro fog power usage effectiveness | 1.4 [57] |
| Best practice access fog power usage effectiveness | 1.5 [57] |
| Number of VMs services | 300 |
| VMs popularity Group | 16%, 5%, 2%, 1%, 0.5% and 0.05% |

Figure 4 shows the power consumption resulting from placing the VMs described above considering the different placement approaches under different users' data rates as explained above. The total power savings achieved under the OC&F approach compared to the single cloud scenario are 5%, 48% and 75%, under 1 Mbps, 10 Mbps and 25 Mbps user data rates, respectively. Figure 5 shows the optimal VMs placement under OC&F approach. Note that the different colors show if a VM of certain popularity is placed in this location or not. Figure 5(a) shows that the cloud in node 2 is selected to serve distributed users of VMs of 1 Mbps users data rate and 0.05% popularity. VMs of a popularity of 0.5% and higher are replicated in every metro fog location except in node 2. Users in node 2 access VMs of 0.5%-16% popularity in the cloud located in node 2 to avoid setting another computing locations in node 2 as we consider on-off power profile. Under the 10 Mbps user data rate, VMs of every popularity group are fully offloaded to metro fog nodes as shown in Fig. 6(b). Under the 25 Mbps user data rate, VMs of ≥ 2% popularity have justified creating a replica copy in each access fog node. Other VMs of 1% popularity and less are offloaded to metro fog nodes. All of the decisions / outcomes described are a function of the relationship between the network and processing parameters considered.

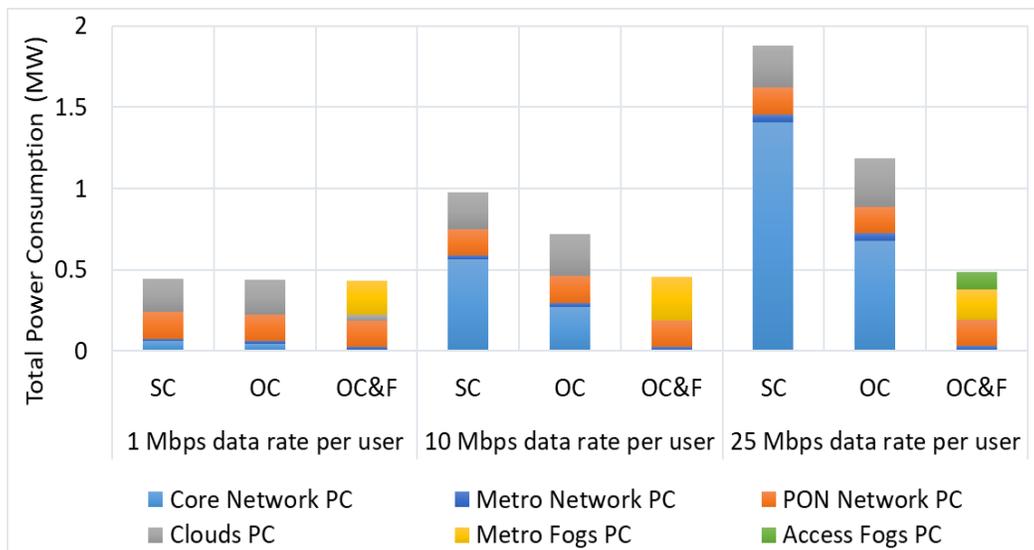

Figure 4: The power consumption of different VMs placement approaches considering VMs of 1% minimum CPU workload.

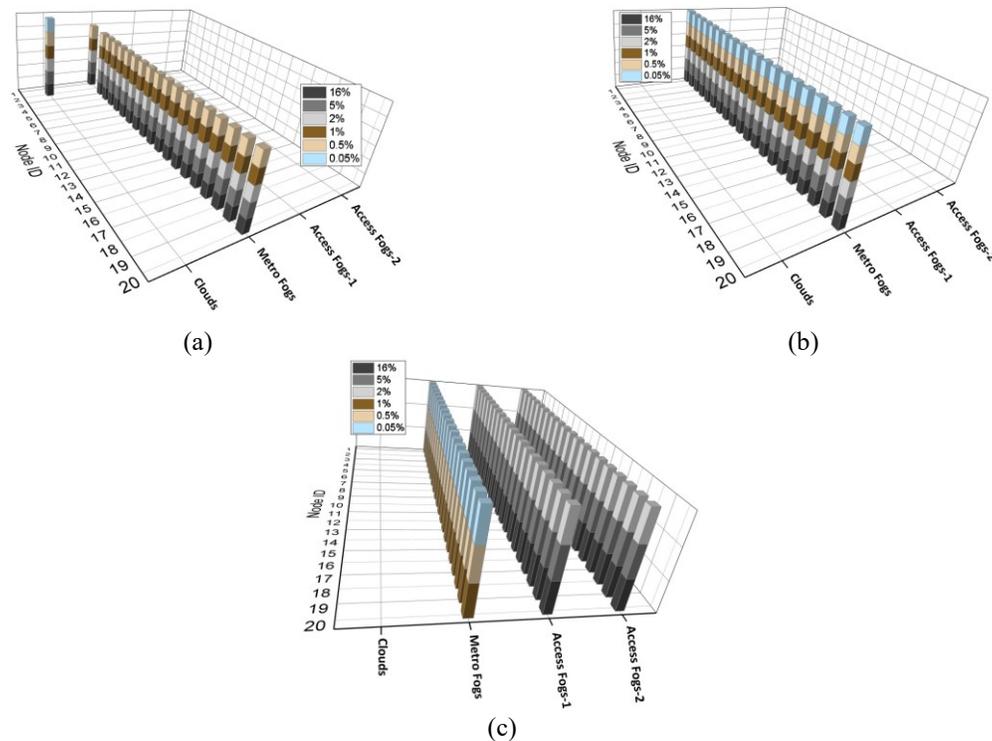

Figure 5: The optimal placement of different VMs popularity groups with 1% minimum CPU workload under the OC&F approach with (a) 1 Mbps data rate per user, (b) 10 Mbps data rate per user and (c) 25 Mbps data rate per user.

IV. Conclusion

Cloud and Fog computing rely on virtual machines (VMs) in order to efficiently use their physical resources. In this paper, we presented a framework for energy efficient VMs placement over a Cloud-Fog architecture. The results show that the decision of offloading VMs to fog nodes is a function of different factors including the workload of the VM, the proximity of fog nodes to users, the VM popularity and the users data rates. Our results showed that optimal VMs placement over a cloud-Fog architecture can save up to 75% of the total power consumption compared to a single cloud scenario.

*Acknowledgments*

The authors would like to acknowledge funding from the Engineering and Physical Sciences Research Council (EPSRC), INTERNET (EP/H040536/1) and STAR (EP/K016873/1) projects. The first author would like to acknowledge the Government of Saudi Arabia and Taibah University for funding his PhD scholarship. All data are provided in full in the results section of this paper.